\documentclass{jnmp}

\usepackage{amsmath}
\usepackage{graphicx}

\JNMPnumberwithin{equation}{section}

\newtheorem{theorem}{Theorem}

\theoremstyle{definition}

\newtheorem*{remark}{Remark} 
\newcommand{\rate}{\overline K}
\newcommand{\nnorm}[1]{\left\|#1\right\|_{\infty,\epsilon}}
\newcommand{\xxx}{\xi}
\newcommand{\yyy}{\zeta}

\begin{document}

%
\renewcommand{\evenhead}{F Leyvraz}
\renewcommand{\oddhead}{Scaling in Irreversible Aggregation Kinetics}

%
\thispagestyle{empty}

\FirstPageHead{**}{**}{20**}{\pageref{firstpage}--\pageref{lastpage}}{Article}

\copyrightnote{2004}{F Leyvraz}

\Name{Rigorous results in the scaling theory of 
irreversible aggregation kinetics}

\label{firstpage}

\Author{Fran\c cois Leyvraz~$^\dag$}

\Address{$^\dag$ Centro de Ciencias F\'\i sicas, University of Mexico, 
Av.~Universidad s/n, Colonia Chamilpa, Cuernavaca, 62131 Morelos, Mexico \\
~~E-mail: leyvraz@fis.unam.mx\\[10pt]
}

\Date{Received Month *, 200*; Revised Month *, 200*; 
Accepted Month *, 200*}

\begin{abstract}
\noindent
The kinetic equations describing irreversible aggregation and
the scaling approach developed to describe them in the limit of large times 
and large sizes are tersely reviewed. Next, a system is considered 
in which aggregates can only react with aggregates of their own size. The 
existence of a scaling solution of the kinetic equations can then be shown 
rigorously in the case in which the total mass of the system is conserved. 
A large number of detailed properties of the solution, previously predicted 
by qualitative arguments, can be shown rigorously as well in this system.
In the case in which gelation occurs, some sketchy rigorous results are 
shown, and numerical evidence for the existence of a scaling solution
is presented. These are the first explicit examples of typical
scaling behaviour for systems exhibiting gelation.
\end{abstract}

%
\section{Introduction}
In many systems one finds that irreversible aggregation of ``clusters'' 
$A(m)$ of mass $m$ occurs and plays an important role. Particular instances 
are aerosol physics, where suspended particles coagulate due to van der 
Waals forces, polymer chemistry and astrophysics (where the clusters may be 
quite varied, going from galaxies to planetary systems). In all these 
systems one is among other things interested in the cluster size 
distribution as a function of time. To obtain it one then relies on kinetic 
equations, derived from the following assumptions: let the
reaction
\begin{equation}
A(m)+A(m^\prime)\mathop{\longrightarrow}_{K(m,m^\prime)}A(m+m^\prime)
\label{eq:1.1}
\end{equation}
occur at a {\em rate\/} $K(m,m^\prime)$, that is, let any two clusters 
of size $m$ and $m^\prime$ react on a time scale given by 
$K(m,m^\prime)^{-1}$. If one further assumes that no spatial correlations 
build up, in other words, that we may use a mean-field model
in which the probability of encounter of two clusters of masses $m_1$ and 
$m_2$
is proportional to 
the product of the concentrations of such clusters, one obtains 
the following kinetic equations for the concentrations $c(m,t)$ of clusters 
of mass $m$:
\begin{gather}
\dot c(m,t)=\frac{1}{2}\int_0^\infty dm_1\,dm_2K(m_1,m_2)\,c(m_1,t)c(m_2,t)
\times\notag\\
\qquad{}\times\left[
\delta(m-m_1-m_2)-\delta(m-m_1)-\delta(m-m_2)
\right].
\label{eq:1.2}
\end{gather}
These are an infinite set of coupled nonlinear ODE's, which represent 
a challenging problem. Few exact solutions are known, which are reviewed 
in \cite{ley03}. 

From the form of the reaction scheme (\ref{eq:1.1}) as well as from the 
equations (\ref{eq:1.2}) it is clear that at the formal level the total 
mass $M$
\begin{equation}
M=\int_0^\infty m\,c(m,t)dm
\label{eq:1.3}
\end{equation}
is conserved. It is well known, however, that this is not generally true. 
If $K(m_1,m_2)$ grows fast enough with the mass, a finite amount of mass 
can escape to infinity in finite time. This phenomenon is known as gelation 
and is found explicitly for the reaction rates $K(m_1,m_2)=m_1m_2$ 
\cite{ley81,McLeoda,McLeodb,zif80}. 

Existence and uniqueness results are known for (\ref{eq:1.2}). Let me state 
two which are representative: White \cite{whi1} has shown that under the 
hypothesis
\begin{equation}
K(m_1,m_2)\leq C(m_1+m_2)
\label{eq:1.4}
\end{equation}
and if the initial condition $c(m,0)$ has finite moments of arbitrary order,
then the kinetic equations (\ref{eq:1.2}) have a unique solution for all 
positive times and this solution preserves mass. A complementary result 
obtained in \cite{ley81} states the
following: under the hypothesis
\begin{gather}
K(m_1,m_2)\leq r(m_1)r(m_2)\nonumber\\
r(m)=o(m),
\label{eq:1.5}
\end{gather}
the kinetic equations (\ref{eq:1.2}) have at least one (positive)
solution for all positive times, which need not conserve mass. This 
was significantly extended in \cite{lau99, lau00}, where it was shown that if
\begin{gather}
K(m_1,m_2)=r(m_1)r(m_2)+\alpha(m_1,m_2)
\label{eq:1.501}\\
\alpha(m_1,m_2)=\alpha(m_2,m_1)\leq Ar(m_1)r(m_2)
\end{gather}
for some positive constant $A$, then a global positive solution exists for 
all times. In this case all requirements of growth on $r(m)$ could be 
dropped, but the rate must now be dominated by a contribution of the 
product type. There also exists a
rigorous result by Jeon \cite{jeo98}, showing that 
no substantial improvement on 
White's theorem is possible: more precisely, for kernels satisfying 
(\ref{eq:1.5}) and additionally
\begin{equation}
K(m_1,m_2)\geq\epsilon(m_1m_2)^\alpha\qquad(\alpha>1/2)
\label{eq:1.6}
\end{equation} 
for some $\epsilon>0$, it is shown that gelation always occurs. 

The rest of this paper is organized as follows: in Sections 2 and 3 I 
tersely review the definitions and predictions of scaling theory for 
non-gelling and gelling systems respectively. In Sectiond 4 and 5 I obtain 
specific results concerning the so-called diagonal kernel in the scaling 
limit. In Ssection 4 I treat the non-gelling case, for which both the 
existence and many qualitative properties of the solution can be rigorously 
established. These results are summarized in Theorem \ref{thm:1}. In 
section 5 I discuss both the gelling case for which I am not 
able to show existence rigorously and the marginal case in 
which the homogeneity degree is one which, although non-gelling by 
White's theorem, lies outside the domain of application of Theorem 
\ref{thm:1}. Existence and qualitative properties of the solution
are summarized in Theorem \ref{thm:2}. I 
present some conclusions in Section 6.
\section{Scaling Theory: The Non-gelling Case}
Beyond the above results on existence of solutions, 
there has always been a great interest in 
knowing the qualitative behaviour of the solutions either in the 
large-time limit or, for gelling systems, for times immediately before the 
singularity. This has been attempted by the so-called scaling description 
of the Smoluchowski equations, to which we now turn. 
\resetfootnoterule
Here we consider the non-gelling case. Scaling theory is then
concerned with the large-size behaviour of the cluster 
size distribution $c(m,t)$. It therefore deals 
exclusively with a limit in which time and size go jointly to infinity 
in a manner to be described shortly.

To proceed, I first need to classify the rates $K(m_1,m_2)$ according to 
the way in which they behave for large values of the arguments. One first 
assumes that they are at the very least asymptotically homogeneous, that 
is, there exists a $\lambda$ such that
\begin{equation}
\rate(x,y)=\lim_{s\to\infty}\left[
s^{-\lambda}K(sx,sy)
\right].
\label{eq:1.10}
\end{equation}
It turns out that only the homogeneous limit $\rate(x,y)$ 
is relevant in the scaling limit. One then needs to classify
homogeneous kernels $\rate(x,y)$ 
further according to the way they behave when $x$ and $y$ are 
very different. One may write $\rate(x,y)$ in the form $x^\lambda k(y/x)$
and, since $\rate(x,y)$ is symmetric, one has
$k(1/\xxx)=\xxx^{-\lambda}k(\xxx)$.
The reaction kernel $\rate(x,y)$ is therefore uniquely specified by the 
function $k(\xxx)$ on the unit interval, which may further be chosen 
arbitrarily.
For simplicity, let us now assume that $k(\xxx)/\xxx^\mu\to1$ 
as $\xxx\to0$,
which defines the exponent $\mu$.
It then follows that $k(\xxx)/\xxx^\nu\to1$ 
as $\xxx\to\infty$, where $\nu=\lambda-\mu$. 
The exponent $\nu$ should satisfy $\nu<1$ for the classical 
existence theorems 
to hold, for which (\ref{eq:1.4}) is required. As stated above,
there exist more general 
theorems \cite{lau99, lau00}, extending to some cases for which $\nu>1$, 
however I shall not consider these here.  

We now turn to the description of the scaling theory for non-gelling 
systems:
it was originally developed by 
Friedlander \cite{fri66,fri00} and later extended by Ernst and van Dongen
\cite{don85}. The underlying idea is the following: if $K(m_1,m_2)$ is 
exactly homogeneous of order $\lambda$, then (\ref{eq:1.2}) may 
have self-similar solutions of the form
\begin{equation}
c(m,t)=Wt^{-2z}\Phi(mt^{-z}),
\label{eq:1.7}
\end{equation}
where $W$ is an unimportant normalization and
the exponent $z$ is given by
$z=1/(1-\lambda)$.
It is a straightforward computation to verify that this is the case if and 
only if the function $\Phi(x)$ satisfies the integral equation: 
\begin{equation}
a^2\Phi(a)=\int_0^a dx\,x\Phi(x)\int_{a-x}^\infty dy\,\rate(x,y)\Phi(y)
\label{eq:1.9}
\end{equation}
and the conditions
\begin{gather}
\int_0^\infty x\Phi(x)dx=1\notag\\
\Phi(x)\geq0.\label{eq:1.9a}
\end{gather}
The problem of showing the existence of self-similar solutions 
is therefore reduced to the corresponding problem for (\ref{eq:1.9}). 
This, however, turns out to be quite difficult. Existence has recently
been shown for three different specific (and rather typical) forms of 
the rate kernel $\rate(x,y)$ in \cite{fou04}, namely:
\begin{subequations}\label{eq:2.5}
\begin{gather}
\rate_1(x,y)=(x^\alpha+y^\alpha)(x^{-\beta}+y^{-\beta})
\qquad(\alpha\in[0,1),\beta\in[0,\infty))\label{eq:2.5a}\\
\rate_2(x,y)=(x^\alpha+y^\alpha)^\beta\qquad(\alpha,\beta\in[0,\infty))
\label{eq:2.5b}\\
\rate_3(x,y)=x^\alpha y^\beta+x^\beta y^\alpha\qquad(\alpha,\beta\in(0,1)).
\label{eq:2.5c}
\end{gather}
\end{subequations}
For these cases Fournier and Lauren\c cot \cite{fou04} have shown existence 
(though not uniqueness) of a 
positive solution to (\ref{eq:1.9}) decaying faster than every power 
for large argument and 
satisfying integrability conditions near the origin corresponding, for 
$\rate_1$ and $\rate_3$, to the results previously conjectured in the 
literature and 
described below. The kernel $\rate_2$ is in a sense, as we shall see 
shortly, marginal and results in this case 
are far more fragmentary. In very closely related work Escobedo {\it et al.}
\cite{esc04} show existence of scaling solutions for a kernel of the same 
type as 
$\rate_3$, but in the parameter range $-1\leq\alpha\leq0\leq\beta\leq1$ 
and $0\leq\lambda<1$. 

Let me describe what is supposed to hold for the behaviour of the 
solution $\Phi(x)$ of (\ref{eq:1.9}) with the boundary condition 
(\ref{eq:1.9a}). For the behaviour near the origin
it has been convincingly argued by power counting arguments
(see for example \cite{don85}) that 
the behaviour is  different according to 
the sign of the exponent $\mu$: if $\mu>0$, then $\Phi(x)$ behaves as 
$x^{-(1+\lambda)}$, whereas if $\mu<0$ it behaves as a stretched exponential:
\begin{equation}
\Phi(x)=\left\{
\begin{array}{ll}
const. \cdot x^{-(1+\lambda)}&(\mu>0)\\
const.\cdot\exp\left(
-const.\cdot x^{-|\mu|}
\right)&(\mu<0),
\end{array}
\right.
\label{eq:1.502}
\end{equation}
where in the second case a power-law behaviour presumably exists as a 
prefactor. 
Finally, if $\mu=0$, the behaviour near the origin is non-universal. This 
case is considerably more complex than the other two, and I shall not 
consider it any further. As an example, we may note that out of the 
kernels in (\ref{eq:2.5a},\ref{eq:2.5b},\ref{eq:2.5c}), $\rate_1$ has 
$\mu=-\beta<0$, $\rate_2$ has $\mu=0$ and $\rate_3$ has 
$\mu=\max(\alpha,\beta)>0$. We therefore see that the three cases are 
represented in the results of \cite{fou04}. As noted there, it could be 
proved that, for $\rate_1$, $\Phi(x)$ decays at the origin faster 
than any power, whereas for $\rate_3$ it could be shown that 
\begin{equation}
\int_0^\infty x^\rho\Phi(x)dx<\infty
\label{eq:1.503}
\end{equation}
for all $\rho>\lambda=\alpha+\beta$, which is a precise form 
of the statement made 
in (\ref{eq:1.502}) and therefore presumably optimal. 

The behaviour for large values of $x$ has been found by 
similar arguments in \cite{don87}. The result is that $\Phi(x)$ decays 
exponentially with a power-law correction, that is
\begin{equation}
\Phi(x)=const.\cdot x^{-\theta}\exp(-const.\cdot x),
\label{eq:1.504}
\end{equation}
where this defines the exponent $\theta$. One then finds 
$\theta=\lambda$ for $\nu<1$,
whereas if $\nu=1$ the behaviour is non-universal and $\theta$ cannot be 
evaluated in general.
\section{Scaling Theory: The Gelling Case}
Let us now pass to the treatment of both the gelling case and that in 
which the degree of homogeneity $\lambda=1$. In the former, it is 
generally assumed, on the basis of numerical work and exactly solved 
models, that 
most of the mass is contained in aggregates of non-singular size. In 
formulae
\begin{equation}
\lim_{M\to\infty}\lim_{t\to t_g}\int_0^Mmc(m,t)dm=1.
\label{eq:1.13}
\end{equation}
Since the distribution $mc(m,t)dm$ is concentrated on small clusters and 
does not, therefore, display the singular behaviour taking place at $t_g$, 
we should look at higher powers, in particular at the distribution
$m^2c(m,t)dm$. Let us assume \cite{ley03} that, after rescaling $m$ by 
an appropriate 
factor, this distribution converges weakly to a distribution 
$x^2\Phi(x)dx$, that is,
\begin{gather}
\lim_{t\to\infty}\left[
M_2(t)^{-1}\int_0^\infty dm\,m^2c(m,t)f\big(m(t_g-t)^{1/\sigma}\big)
\right]=\int_0^\infty dx\,x^2\Phi(x)f(x)\notag\\
M_2(t)=\int_0^\infty m^2c(m,t)dm.
\label{eq:1.15}
\end{gather}
Here the rate at which the typical size diverges is given traditionally
by the exponent $-1/\sigma$. 
In order for these equations to be closed, however, we need further 
knowledge concerning the behaviour of $M_2(t)$ near $t_g$. This can be 
obtained by the following argument. It is clear that for gelation to occur 
the second moment must diverge at $t_g$. The concentrations $c(m,t_g)$ have 
therefore some kind of power-law behaviour which we may denote by 
$m^{-\tau}$. If we now take $M_2(t)$ to be approximated by
\begin{equation}
\int_0^{(t_g-t)^{-1/\sigma}}m^2c(m,t)dm=(t_g-t)^{-(3-\tau)/\sigma}
\label{eq:1.16}
\end{equation}
and if we additionally assume that the exponent $\tau$ characterizing 
particles of fixed size $m$ as $t\to t_g$ coincides with a power-law 
singularity at the origin of the scaling regime, in other words, if we 
assume that particles of size very small 
with respect to the typical size have the same 
power-law singularity as particles of fixed size when that size goes to 
infinity, then we are led to the additional boundary condition
\begin{equation}
\Phi(x)=const.\cdot x^{-\tau}\qquad(x\to0).
\label{eq:1.17}
\end{equation}
These assumptions from now on lead in a rather straightforward way to the 
following extension of (\ref{eq:1.9}) for gelling systems: 
\begin{gather}
\int dx\left[
(3-\tau)f(x)+xf^\prime(x)
\right]x^2\Phi(x)
=
\int dx\,dy\,\rate(x,y)x\Phi(x)\Phi(y)\times\notag\\
\qquad\times\left[
(x+y)f(x+y)-xf(x)
\right]
\label{eq:1.18}
\end{gather}
for all continuously differentiable functions $f(x)$. The problem is 
therefore reduced to a kind of non-linear eigenvalue problem: one must 
find a value of $\tau$ such that there exists a solution $\Phi(x)$
of (\ref{eq:1.18}) satisfying the boundary condition (\ref{eq:1.17}) and 
decaying sufficiently fast at infinity. We shall show in an explicit 
example, namely the so-called diagonal kernel $\rate(x,y)$ given by 
$x^{\lambda+1}\delta(x-y)$, how the problem may be solved at least in 
practice and give strong numerical arguments for the existence and 
uniqueness of a solution. 

For the exponent 
$\sigma$ characterizing the divergence of the typical size, one obtains
\cite{ley03}
\begin{equation}
\sigma=1+\lambda-\tau,
\label{eq:1.19}
\end{equation}
in stark contrast with the non-gelling case, in which the corresponding 
exponent $z$ could be expressed directly in terms of the known value 
$\lambda$. Here, on the other hand, (\ref{eq:1.19}) only yields a relation 
between two equally unknown quantities. There does exist \cite{ley03}
an interesting 
relation between $\sigma$ and the correction to scaling exponent $\Delta$ 
of $\Phi(x)$ defined by
\begin{equation}
\Phi(x)=const.\cdot x^{-\tau}\left[1+O(x^\Delta)\right].
\label{eq:1.20}
\end{equation}
It is then conjectured that (see \cite{ley03} for details) that
\begin{equation}
\Delta=\sigma=1+\lambda-\tau.
\label{eq:1.21}
\end{equation}
In this paper we shall show rigorously for the case of the diagonal kernel
that, if a solution to the non-linear eigenvalue problem described above 
actually exists, then the above relation holds. 
\section{The Diagonal Kernel: The Non-gelling Case}
In the following we study in greater detail the so-called diagonal kernel 
given by $\rate(x,y)=x^{\lambda+1}\delta(x-y)$
Physically, this means that only aggregates of the same size react with 
each other. This might at first be viewed with suspicion: for example, if 
one starts with initial conditions of the type
$c(m,0)=\delta(m-m_0)$
then a very singular solution arises, in which only sizes 
of the form $2^km_0$ occur. 
However, we may start from less singular initial conditions and should also 
remember that in the scaling limit any kernel which only allows reactions 
between nearby sizes would be described asymptotically by this kernel. 

It follows from White's theorem \cite{whi1} that $\lambda\leq1$ is 
non-gelling, whereas a specific study of this particular model 
\cite{ley83,buf89} shows that the system undergoes gelation for 
$\lambda>1$\footnote{Note that this does not follow directly from Jeon's 
more general result in \cite{jeo98}.}.
In this section we only consider $\lambda<1$, as the case $\lambda=1$ turns 
out to be highly singular and best treated by methods akin to those used 
for the gelling case. One can 
then be quite explicit about the properties of the solution of 
(\ref{eq:1.9}): 
\begin{theorem}
\label{thm:1}
Let the reaction kernel $\rate(x,y)$ be given by
$x^{\lambda+1}\delta(x-y)$
where $\lambda<1$ and let $\Delta$ be the unique positive solution of the 
transcendental equation
\begin{equation}
\frac{1+\Delta}{2}=\frac{1-2^{(\lambda-1)(\Delta+1)}}{1-2^{\lambda-1}}.
\label{eq:2.14}
\end{equation}
Then there is an $\epsilon>0$ and a solution $\Phi(x)$á
of (\ref{eq:1.9}) such that
\begin{equation}
\Phi(x)=x^{-(1+\lambda)}\left[
\frac{1-\lambda}{1-2^{\lambda-1}}-cx^\Delta+O
\left(x^{\Delta+\epsilon}\right)
\right].
\label{eq:2.15}
\end{equation}
Here $c$ is an arbitrary positive constant, which can be varied
so as to set the 
normalization (\ref{eq:1.9a}). Furthermore, if $\Phi(x)$ is of the form 
(\ref{eq:2.15}), then it is uniquely determined by $c$. Finally, 
$x^{1+\lambda}\Phi(x)$ is monotonically decreasing  on $[0,\infty)$
as well as bounded by an exponential function $C\exp(-ax)$ for some 
positive constants $C$ and $a$.
It is also $C^\infty$ on $[0,\infty)$ in the variable 
$x^\Delta$.
\end{theorem}
\begin{remark}
Note that the Theorem makes no claim concerning uniqueness of $\Phi(x)$ 
if $\Phi(x)$ does not have the form (\ref{eq:2.15}). In particular, 
solutions having divergent second moments, as found in the case of the 
additive kernel by Menon and Pego \cite{men03} are not excluded for this 
kernel. It would certainly be of great interest to know whether such solutions 
exist. 
Finally, $x^{1+\lambda}\Phi(x)$ can be 
shown to be analytic in $x^\Delta$ in a neighbourhood of the origin.
I believe, but cannot prove, that this
is actually true for the entire positive real axis. 
\end{remark}
\begin{proof}
Let us first write (\ref{eq:1.9}) out explicitly in the case 
of the diagonal kernel. One finds
\begin{equation}
x^2\Phi(x)=\int_{x/2}^xdy\,y^{\lambda+2}\Phi(y)^2,
\label{eq:3.1}
\end{equation}
which can easily be rewritten as a ``non-local'' differential equation:
\begin{equation}
\frac{d}{dx}\left[
x^2\Phi(x)
\right]=x^{\lambda+2}\Phi(x)^2-\frac{1}{2}\left(\frac{x}{2}\right)^{\lambda+2}
\Phi\left(
\frac{x}{2}
\right)^2.
\label{eq:3.2}
\end{equation}
We now perform the following change of variables
\begin{eqnarray}
\psi&=&x^{1+\lambda}\Phi\nonumber\\
y&=&\frac{x^{1-\lambda}}{1-\lambda},
\label{eq:3.3}
\end{eqnarray}
which casts (\ref{eq:3.2}) in the form
\begin{equation}
(1-\lambda)\frac{d}{dy}\left[
y\psi(y)
\right]=\psi(y)^2-2^{\lambda-1}\psi(2^{\lambda-1}y)^2.
\label{eq:3.4}
\end{equation}
It is clear that the only way in which such a function can tend to a 
constant value $\psi_0$ at the origin arises if
\begin{equation}
\psi_0=\frac{1-\lambda}{1-2^{\lambda-1}}.
\label{eq:3.5}
\end{equation}
We therefore begin by searching for solutions of the type
\begin{equation}
\psi(y)=\psi_0+\phi(y).
\label{eq:3.6}
\end{equation}
$\phi(y)$ then satisfies the equation
\begin{equation}
(1-\lambda)\left[
y\phi(y)\right]^\prime=2\psi_0\left[
\phi(y)-2^{\lambda-1}\phi(2^{\lambda-1}y)
\right]+\phi(y)^2-2^{\lambda-1}\phi(2^{\lambda-1}y)^2,
\label{eq:3.7}
\end{equation}
which always has the trivial solution $\phi(y)=0$ corresponding to 
the solution of (\ref{eq:3.1}) given by $\Phi(x)=\psi_0x^{-(1+\lambda)}$. 
The existence of such a formal solution, which cannot satisfy the 
normalization condition (\ref{eq:1.9a}) is a well-known fact for all 
kernels of type I. We should therefore look for ways of violating 
uniqueness for the solution of (\ref{eq:3.7}), which can clearly only 
happen near the origin. 

If we wish to construct a solution starting at $\psi_0$ and having the 
form (\ref{eq:3.6}), then $\phi(y)$ must initially be small. This suggests 
looking at the linearization of (\ref{eq:3.7}), which
can be solved by an appropriate power law $y^\Delta$. 
One then obtains the transcendental equation
\begin{equation}
\frac{1+\Delta}{2}=\frac{1-2^{(\lambda-1)(\Delta+1)}}{1-2^{\lambda-1}}=F(\Delta).
\label{eq:3.9}
\end{equation}
Here the second equality {\em defines\/} $F(\Delta)$.  
The solution is found to exist and to be unique: indeed, one finds 
$F(0)=1>1/2$, but $F(\Delta)$ becomes less than $(1+\Delta)/2$ as 
$\Delta\to\pm\infty$, so (\ref{eq:3.9}) must have both
a positive and a negative 
root. But since $F(\Delta)$ is concave, there are no more than two 
roots, so that the positive solution is unique as stated. Further, since 
$F(\Delta)>1$ for all $\Delta>0$, it follows from (\ref{eq:3.9}) that 
$\Delta>1$.  

It now remains to show that a solution $\phi(y)$ of the type described 
above indeed exists in a sufficiently small neighbourhood of the origin. To 
this end we introduce
\begin{equation}
\phi(y)=y^\Delta \left[-c+h(y)\right],
\label{eq:3.10}
\end{equation}
where $c$ is an arbitrary constant, which we shall later take to be 
positive, but its sign is of no importance for the {\em local\/} result I 
am now proving.
Using the transcendental equation (\ref{eq:3.9}) satisfied by $\Delta$, we 
may now recast the equation (\ref{eq:3.1}) as follows
\begin{equation}
h(y)=\frac{1}{(1-\lambda)y^{1+\Delta}}\int_{2^{\lambda-1}y}^y
dw\,w^\Delta\left\{
2\psi_0h(w)+w^\Delta\left[
-c+h(w)
\right]^2
\right\}.
\label{eq:3.12}
\end{equation}
To show that this fixed point problem has a unique solution, it is 
sufficient to find a closed subset of a Banach space which is left invariant 
by the operator
\begin{equation}
T[h](y)=\frac{1}{(1-\lambda)y^{1+\Delta}}\int_{2^{\lambda-1}y}^y
dw\,w^\Delta\left\{
2\psi_0h(w)+w^\Delta\left[
-c+h(w)
\right]^2
\right\}
\label{eq:3.13}
\end{equation}
and in which $T$ is a contraction. Define $C_{\infty,\epsilon}(y)$ 
to be the space of continuous functions vanishing at the origin 
and so that there exists a $C$ such that
\begin{equation}
\yyy^{-\epsilon}g(\yyy)\leq C\qquad(0\leq \yyy\leq y).
\label{eq:3.14}
\end{equation}
If one then defines
\begin{equation}
\|h\|_{\infty,\epsilon}=\sup_{0\leq w\leq y}\left|w^{-\epsilon}h(w)\right|,
\label{eq:3.15}
\end{equation}
it is easy to see that the operator $T$ maps $C_{\infty,\epsilon}(y)$
on itself as long as $\epsilon\leq\Delta$. It is further straightforward 
to show that 
\begin{equation}
\nnorm{T[h]}\leq C(\epsilon)\nnorm{h}+a(y)\nnorm{h}^2+b(y),
\label{eq:3.151}
\end{equation}
where $C(\epsilon)<1$ whereas $a(y)$ and $b(y)$ go to zero as $y\to0$. From 
this follows that $T$ leaves a ball of appropriately small radius $R$ 
invariant if $\epsilon$ is fixed and $y$ is small enough. 

To show that $T$ is in fact a 
contraction on such a ball for $y$ sufficiently small, one computes 
the norm of
\begin{gather}
T[h_1]-T[h_2]=
\frac{1}{(1-\lambda)y^{1+\Delta}}\int_{2^{\lambda-1}y}^y
dw\,w^\Delta\big\{
2\psi_0\left[h_1(w)-h_2(w)
\right]+\notag\\
\qquad+w^\Delta\left[
h_1(w)-h_2(w)
\right]\left[
h_1(w)+h_2(w)-2c
\right]
\big\}.
\label{eq:3.16}
\end{gather}
For the first term one shows
\begin{equation}
\frac{1}{(1-\lambda)y^{1+\Delta+\epsilon}}\int_{2^{\lambda-1}y}^y
dw\,w^\Delta
2\psi_0\left[h_1(w)-h_2(w)
\right]\leq\frac{2\psi_0}{1-\lambda}\int_{2^{\lambda-1}}^1
dw\,w^{\Delta+\epsilon}\cdot\nnorm{h_1-h_2}.
\label{eq:3.17}
\end{equation}
It then readily follows from the equation (\ref{eq:3.9}) 
defining $\Delta$ that the 
numerical constant before $\nnorm{h}$ is strictly less than one if 
$\epsilon>0$. For the second term, on the other hand, one finds, as is 
easily checked, that it is bounded by
\begin{equation}
C(y)(\nnorm{h_1}+\nnorm{h_2})\nnorm{h_1-h_2}
\label{eq:3.171}
\end{equation}
where $C(y)$ goes to zero 
as $y$ does. Since we have restricted ourseleves to a ball of given radius 
$R$, the prefactor $\nnorm{h_1}+\nnorm{h_2}$ is bounded by $2R$, so that the 
entire mapping $T$ is indeed a contraction for $y$ small enough. 

To show global existence and decay to zero, we must first choose $c>0$. 
From this immediately follows that there is a neighbourhood of zero in 
which the solution $\psi(y)$ is monotonically decreasing. One now shows 
using (\ref{eq:3.4}) that this property never disappears for any positive 
value of $y$. Indeed, if $\psi(y)$ were to increase for some value of $y$, 
it would have had to pass through a minimum at some value of $y$. Let $y_0$ 
be the first non-zero value of $y$ for which $\psi^\prime(y)=0$. One then 
finds
\begin{gather}
0=\psi(y_0)^2-(1-\lambda)\psi(y_0)-2^{\lambda-1}\psi(2^{\lambda-1}y_0)^2
\notag\\
\qquad<(1-2^{\lambda-1})\psi(y_0)^2-(1-\lambda)\psi(y_0)
\label{eq:3.18}\\
\qquad<\psi(y_0)\left[
(1-2^{\lambda-1})\psi_0-(1-\lambda)
\right]=0,\notag
\end{gather}
which is absurd. Here the first inequality follows from the monotonicity 
of $\psi(y)$ before $y_0$, and so does the second, since $\psi_0$ is the 
initial value of $\psi(y)$. Both are strict since $\psi^\prime(y)\neq0$
for $y<y_0$. 

Once the local solution near the origin has been found, however, the 
equation (\ref{eq:3.4}) can be considered as an ODE and the standard 
existence and uniqueness results apply: indeed, the function 
$\psi(2^{\lambda-1}y)$ is {\em already known\/} at $y$. In other words, 
if $\psi_n(y)$ is defined to be $\psi(y)$ restricted to the interval $I_n$ 
given by $[2^{(1-\lambda)(n-1)},2^{(1-\lambda)n}]$, then equation 
(\ref{eq:3.4}) can be viewed as an ODE connecting $\psi_n^\prime$ 
and $\psi_{n-1}$. This then allows a recursive application of the usual 
existence and uniqueness results, since there is, by the local result just 
proved, a unique solution for $n$ sufficiently negative. 
Furthermore, the 
integral form of the equation for $\psi(y)$, which is given by
\begin{equation}
(1-\lambda)y\psi(y)=\int_{2^{\lambda-1}y}^ydw\,\psi(w)^2
\label{eq:3.19}
\end{equation}
guarantees the positivity of $\psi(y)$ wherever it exists. With these 
remarks, standard theorems on ordinary differential equations yield global 
existence of $\psi(y)$ on the positive real axis. 

Proving that $\psi(y)$ decays to zero exponentially fast as $y\to\infty$ is 
now straightforward: indeed, from (\ref{eq:3.19}) and the monotonicity 
properties of $\psi(y)$ follows immediately
\begin{equation}
\psi(y)\leq\frac{1-2^{\lambda-1}}{1-\lambda}\psi(2^{\lambda-1}y)^2
=\psi_0^{-1}\psi(2^{\lambda-1}y)^2,
\label{eq:3.20}
\end{equation}
from which one obtains exponential decay in the following way: since 
$\psi(y)<\psi_0$ for $y>0$, by applying this identity repeatedly one 
eventually shows that there is an interval $I_{n_0}$ such that $\psi(y)\leq 
M<1$ on it. Now, 
if $x\in 
I_n$, it follows that the inequality (\ref{eq:3.20}) can be applied 
iteratively $n-n_0$ times to yield
\begin{equation}
\psi(y)\leq\psi_0^{-(n-n_0)}\psi(2^{(n-n_0)(\lambda-1)}y)^{2^{n-n_0}}\leq 
\psi_0^{-(n-n_0)}\,M^{2^{n-n_0}},
\label{eq:3.21}
\end{equation}
from which exponential decay in $y$ follows, since $y\in I_n$ means that
$(1-\lambda)(n-1)\ln 2\leq \ln y\leq (1-\lambda)n\ln 2$. Note that the 
prefactor increasing exponentially in $n$ corresponds only to a power law 
in $y$ and is therefore a subdominant correction. 

I mention here an easier way of proving a local existence result, but 
which may be difficult to extend to an equally strong uniqueness result: 
one can make the Ansatz
\begin{equation}
\psi(y)=\psi_0+\sum_{n=1}^\infty c_n y^{n\Delta}
\label{eq:3.22}
\end{equation}
and plug this in (\ref{eq:3.4}) keeping at first $\Delta$ a free parameter. 
A recursion for the $c_n$ is easily derived, which 
is, however, in general inconsistent for $c_1$. The consistency condition is, 
of course, the transcendental equation (\ref{eq:3.9}) for $\Delta$. $c_1$ 
can then be chosen arbitrarily, thereby determining the $c_n$ uniquely. 
Convergence is also readily verified by 
inspection of the recurrence. 
This result shows as an added bonus that $\psi(y)$ is analytic in 
$y^\Delta$ around the origin. 

From the result shown we have therefore proved that for each negative 
value of $c_1$ there exists a solution of the original equation 
(\ref{eq:3.1}) of the form
\begin{equation}
\Phi(x)=x^{-(1+\lambda)}\left[
\psi_0+\sum_{n=1}c_nx^{n\Delta}
\right]
\label{eq:3.23}
\end{equation}
for small values of $x$ and that this function exists on the 
entire positive real axis as a positive function which decays 
exponentially to zero. It therefore satisfies the scaling equation and all 
the boundary conditions imposed upon it, since it is easy to see that the 
normalization to one of the first moment of $\Phi(x)$ can be obtained 
through an appropriate choice of $c_1$, as the first moment is 
finite for functions of the type described by (\ref{eq:3.23}).
\end{proof}
\section{The Diagonal Kernel: The Gelling Case}
Let us now consider the same diagonal kernel $x^{\lambda+1}\delta(x-y)$ 
but now for 
the case $\lambda\geq1$. We first consider $\lambda>1$ and then say a few 
words concerning the case for which $\lambda=1$, which does not display 
gelation but for which the ordinary scaling ansatz described in the previous 
section is nevertheless inapplicable. 

The relevant equation (\ref{eq:1.18}) reduces for the diagonal case to
\begin{equation}
\frac{d}{dx}\left[
x^3\Phi(x)
\right]-
(3-\tau)x^2\Phi(x)=x^{\lambda+3}\Phi(x)^2-
2^{-(\lambda+3)}x^{\lambda+3}\Phi(x/2)^2,
\label{eq:5.1}
\end{equation}
where $\Phi(x)$ should satisfy the boundary condition at zero
\begin{equation}
\Phi(x)=const.\cdot x^{-\tau}\qquad(x\to0)
\label{eq:5.2}
\end{equation}
as well as a condition of rapid decay as $x\to\infty$.
One now introduces new variables
\begin{gather}
\psi=x^\tau\Phi(x)\notag\\
\yyy=\frac{x^{1+\lambda-\tau}}{1+\lambda-\tau}.
\label{eq:5.3}
\end{gather}
Note that in our new variables the 
condition (\ref{eq:5.2}) translates into the requirement that $\psi$ 
be regular at the origin. We have also made the implicit 
assumption that $1+\lambda-\tau>0$: this is quite unproblematic, 
since through the relation 
(\ref{eq:1.19}) determining $\sigma$ it follows that this must always 
hold if we wish to describe a gelling system. 

The relation (\ref{eq:5.1}) now transforms to
\begin{equation}
\frac{d\psi}{d\yyy}=\psi(\yyy)^2-2^{2\tau-\lambda-3}\psi\big(
2^{\tau-\lambda-1}\yyy
\big)^2.
\label{eq:5.4}
\end{equation}
Local existence is here easy to show: one makes an ansatz in power series
\begin{equation}
\psi(\yyy)=\sum_{k=0}^\infty a_k\yyy^k.
\label{eq:5.5}
\end{equation}
It is then 
straightforward to determine a recursion for the $a_k$ and to prove that 
it converges in a finite interval. 
No inconsistencies arise in this recursion, so we may simply 
set $a_0=1$. It follows that
\begin{equation}
a_1=1-2^{2\tau-\lambda-3}.
\label{eq:5.999}
\end{equation}
Let us then always limit ourselves to the case $\tau>(\lambda+3)/2$, so
as to have an initially monotonically decaying function. Note here that 
this shows the claim made in the text, that the correction to scaling 
exponent of $\Phi(x)$ is $1+\lambda-\tau$: indeed, we have shown that 
around $\yyy=0$, $\psi(\yyy)$ is analytic, which together with the definition 
of $\yyy$ in (\ref{eq:5.3}) shows the result. In fact, the result obtained is 
stronger: it states that $\Phi(x)$ is of the form
\begin{equation}
\Phi(x)=x^{-\tau}f\left(x^{1+\lambda-\tau}\right),
\label{eq:5.61}
\end{equation}
where $f(\yyy)$ is a function analytic around zero. 

Since $\tau>(\lambda+3)/2$ one can show that
this leads, as in the preceding section, to global existence unless the 
solution becomes negative. It is a crucial difference with the previous 
case that this can actually happen for given values of $\tau$, as we shall 
see shortly. One proceeds as before: let $\yyy_0$ be the 
first value of $\yyy$ for which $\psi^\prime(\yyy)=0$. Since $\psi(\yyy)$ is 
monotonically decreasing for $\yyy<\yyy_0$, one finds using (\ref{eq:5.4}) 
that $\psi^\prime(\yyy_0)<0$, which is clearly absurd. 

\begin{figure}
\include{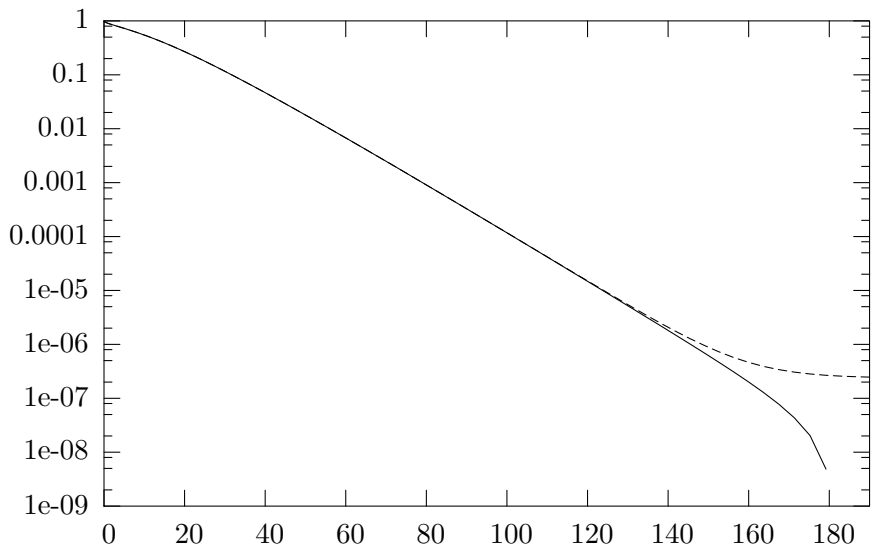}
\caption{Two solutions of (\ref{eq:5.4}) for $\lambda=2$
corresponding to two nearby 
values of $\tau$ plotted as a function of $x$, that is 
$\yyy^{1/(1+\lambda-\tau})$: for the lower curve $\tau=2.5542887$ whereas 
for the 
upper one $\tau=2.5542888$. Note how one crosses over to negative values, 
whereas the other seems to saturate. Note also the rapid decay common to 
both curves, which suggests the existence of an exponentially decaying 
solution of (\ref{eq:5.4}) for some critical $\tau$.}
\label{fig:1}
\end{figure}

Performing a detailed qualitative analysis of (\ref{eq:5.4}) is much more 
difficult than the corresponding task in the non-gelling case. I shall 
therefore appeal to numerical evidence, which seems to me quite compelling. 
Let us disregard considerations of uniqueness and look for a solution in 
the region $\tau>(\lambda+3)/2$, for which I know that the solution is 
always decreasing. $\psi(\yyy)$ therefore either crosses to negative values 
at some value of $\yyy$, in which case it is surely not a physically relevant 
solution, or else it exists as a positive function for all positive values 
of $\yyy$. In this latter case it is also easy to check that it must go to 
zero, as it cannot saturate to a constant non-zero value unless 
$\tau=(\lambda+3)/2$. 
\begin{figure}
\include{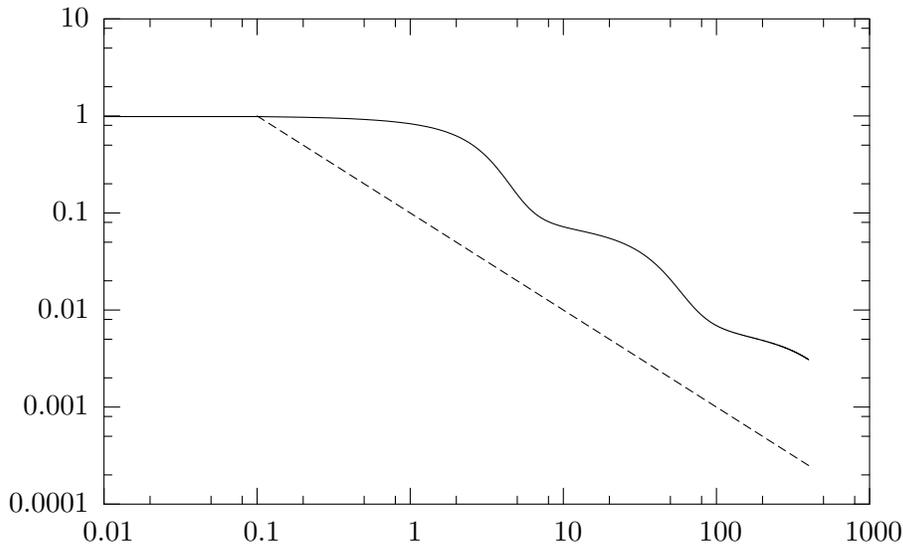}
\caption{A solution of (\ref{eq:5.4}) corresponding to $\lambda=2$
and $\tau=2.6$ plotted as a function of $\yyy$: the dotted curve corresponds 
to $1/\yyy$ behaviour. Note the 
overall parallelism as well as the logarithmic oscillations of the 
function. The occurrence of such oscillations is intriguing, as they have 
been tentatively
reported by Lee \cite{lee01} for the scaling function in non-gelling 
systems of type I. The curve shown here, 
on the other hand, is an inadmissible solution for 
a gelling system.}
\label{fig:2}
\end{figure}
If $\tau$ is barely larger than $(\lambda+3)/2$, it is relatively easy to 
see that $\psi(\yyy)$ must cross to negative values and 
numerical work amply confirms this 
expectation. This is found by comparing (\ref{eq:5.4}) with
\begin{equation}
\frac{df}{d\yyy}=f(\yyy)^2-(1+\epsilon)f(0)^2,
\label{eq:5.610}
\end{equation}
which is close, as for such values of $\tau$ $\psi(\yyy)$ is nearly 
constant. It then suffices to show that, for small values of $\epsilon$, 
the solution of (\ref{eq:5.610}) crosses the negative axis for values of 
$\yyy$ at which the value $\psi(2^{\tau-\lambda-1}\yyy)$ is still close to 
its original value.
Another asymptotic behaviour which is consistent with 
(\ref{eq:5.4}) is a $\yyy^{-1}$ decay. Such a decay can in fact be shown to 
occur in the case $\tau=\lambda+1$, since (\ref{eq:5.4}) can then be 
solved exactly and is found to have this behaviour. 
Again, as shown in Fig.~\ref{fig:2}
for the case $\lambda=2$ and $\tau=2.6$, such 
a decay is indeed observed for values of $\tau$ somewhat larger than those
for which a crossing to negative values was observed. 
What we want, however, is neither: negative 
values are unphysical and the $\yyy^{-1}$ is too slow to be acceptable. One
needs a function having a rapid decay for large $\yyy$. 
We may therefore 
ask whether between the two regimes there is a (conceivably unique) value 
of $\tau$ for which this occurs. In fact, we anticipate exponential 
decay in $x$ and hence a corresponding streched exponential decay in $\yyy$. 
Fig.~\ref{fig:1} gives a striking confirmation of this suggestion: 
it plots for 
$\lambda=2$ two quite nearby values of $\tau$. For the lesser, we see a 
crossing to negative values, for the higher, a saturation apparently to a 
constant (but theory 
states this to be impossible, so it is presumably a $\yyy^{-1}$ behaviour), 
but both exhibit over a very large range exactly the expected exponential 
decay: for clarity I have plotted $\psi(\yyy)$ as a function of 
$x$ on a semilogarithmic scale, for which the expected result is 
a straight line.

Let us end by some remarks concerning the limiting case $\lambda=1$. If we 
refer to the results of the previous section for the non-gelling case, we 
find that they do not apply, as all equations become highly singular and 
the proofs all fail. The reason for this is to be found in the fact that 
if Theorem \ref{thm:1} were valid,
$\Phi(x)$ would have an $x^{-2}$ singularity at the origin, 
thereby precluding the use of the scaling hypothesis (\ref{eq:1.9}). Indeed, 
for (\ref{eq:1.9}) to hold, the normalization condition (\ref{eq:1.9a}) 
is necessary, and this is contradicted by the existence of such a 
singularity. We therefore treat the subject as if it were a gelling case 
and study the second moment as in (\ref{eq:1.15}). This leads \cite{ley03}
to a typical size $s(t)$ of the form
\begin{equation}
s(t)=const.\cdot\exp(K\sqrt t)
\label{eq:5.8}
\end{equation}
and a function $\Phi(x)$ satisfying 
\begin{equation}
\frac{d}{dx}\left[
x^3\Phi(x)
\right]-
x^2\Phi(x)=x^4\Phi(x)^2-
(x/2)^4\Phi(x/2)^2,
\label{eq:5.9}
\end{equation}
which is merely (\ref{eq:5.1}) in the special case in which $\lambda=1$ 
and $\tau=2$. 

We now state our results for this case:
\begin{theorem}
\label{thm:2}
For each $a_1<0$ (\ref{eq:5.9}) has a unique solution of the form
\begin{equation}
\Phi(x)=x^{-2}\left[
1+\sum_{k=1}^\infty a_kx^k
\right].
\label{eq:5.91}
\end{equation}
$x^2\Phi(x)$ is monotonically decreasing on the positive real axis
and bounded by an exponential function $C\exp(-ax)$ for some positive 
constants $C$ and $a$. It is 
also $C_\infty$ in $[0,\infty)$. 
\end{theorem}
\begin{remark}
Note that there are other solutions to (\ref{eq:5.9}) which are not of the 
form (\ref{eq:5.91}). In particular, the same proof as below shows that 
there are solutions of the form
\begin{equation}
\Phi(x)=x^{-2}\left[
\psi_0+\sum_{k=1}^\infty a_kx^{k\Delta}
\right]
\label{eq:5.92}
\end{equation}
for all $\psi_0>0$
where $\Delta$ is the unique positive solution of
\begin{equation}
\frac{\Delta}{2(1-2^{-\Delta})}=\psi_0.
\label{eq:5.93}
\end{equation}
It is readily seen, however, that these solutions bring no new scaling 
solutions, since the exponent $\Delta$ can be absorbed in the constant $K$ 
of (\ref{eq:5.8}). I also am not sure whether solutions with power-law 
tails having a form different from (\ref{eq:5.92}) could exist. 
\end{remark}
\begin{proof}
The change of independent variable in (\ref{eq:5.3}) now 
becomes singular, so we merely change to
\begin{equation}
\psi=x^2\Phi(x).
\label{eq:5.10}
\end{equation}
Note that we still have the boundary condition $\psi(x)$ regular at the 
origin, corresponding to (\ref{eq:5.2}) for $\tau=2$. This then yields
\begin{equation}
x\frac{d\psi}{dx}=\psi(x)^2-\psi(x/2)^2.
\label{eq:5.11}
\end{equation}
One finds a solution as follows: the ansatz
\begin{equation}
\psi(x)=1+\sum_{k=1}^\infty a_kx^k
\label{eq:5.12}
\end{equation}
leads, as is readily verified, to a recurrence for the $a_k$ 
which is consistent throughout and for 
which $a_1$ can be chosen arbitrarily, after which all $a_k$ are 
uniquely determined. Convergence of the series (\ref{eq:5.12}) 
in a finite interval around zero is also 
easily checked. Again choose $a_1$ to be negative. It 
follows immediately that $\psi(x)$ is monotonically decreasing 
in a neighbourhood of the origin and hence, by an argument entirely 
similar to that presented in Section 4, wherever it 
exists on the positive real axis. The equation can also be rewritten in 
integral form as
\begin{equation}
\psi(x)=\int_{x/2}^x\frac{dy}{y}\psi(y)^2,
\label{eq:5.13}
\end{equation}
which shows that $\psi(x)$ is always positive on the positive real axis. 
This leads, using the same approaches as previously, to a global existence 
result for the solution of (\ref{eq:5.9}). From (\ref{eq:5.13}) one also 
immediately derives the inequality
\begin{equation}
\psi(x)\leq\ln2\,\psi(x/2)^2
\label{eq:5.14}
\end{equation}
from which, as in the previous section, exponential decay 
of $\psi(x)$ is readily
deduced.
\end{proof}
\section{Conclusions and Outlook}
Summarizing, we have shown that the equations describing  the scaling 
function for the Smoluchowski equations describing irreversible 
aggregation have a unique solution within a given family of physically 
reasonable solutions when the reaction rates are of such a nature as only 
to allow reactions between aggregates of identical sizes. Since this 
kernel is dominated by diagonal reactions, it belongs to the so-called 
type I kernels. For these the very existence of a solution to the scaling 
equation (\ref{eq:1.9}) has sometimes been doubted, both on theoretical and 
numerical grounds: theoretically, because of the existence of a 
non-normalizable solution of the type $x^{-(1+\lambda)}$, which indicates 
that the constraint of fast decay at infinity must play an essential role 
in determining the solution. Numerical work on some kernels of type I also 
led to doubts concerning scaling \cite{kri95}, though these were later 
contradicted by \cite{lee01}, who showed that the previous finding may have 
been due to non-monotonic behaviour of the scaling function for the rates 
under consideration. For the diagonal case under study, I have not only 
rigorously shown existence in the non-gelling case, but also 
monotonicity.
For the gelling case, on the other hand, it was suggested by numerical 
evidence that whenever $\tau>(\lambda+3)/2$, there are only two possible
behaviours in general, namely becoming negative or decaying 
algebraically to zero from above. 
Just at the border between these two unphysical
behaviours, numerical 
evidence again strongly suggests that a positive and exponentially decaying 
solution exists. This then allows a determination of the exponent
$\tau$. As stated already in \cite{ley03}, it is not to be expected that 
the ``standard'' value $\tau=(\lambda+3)/2$ 
broadly quoted in  the literature
should be generally valid. In 
fact, it follows from the considerations made in this paper that no 
solutions of this type can exist for the diagonal kernel. 
For the limiting case $\lambda=1$, we have shown rigorously
the existence of a solution of the scaling equation (\ref{eq:1.18}) 
derived from an 
assumed weak convergence of the (appropriately rescaled)
measure $m^2c(m,t)dm$ to a limiting 
distribution $x^2\Phi(x)dx$. These are related to the modified 
scaling 
ansatz introduced by van Dongen \cite{don88}. No such explicit solutions 
have, to my knowledge, been derived before.

Finally, it must be remembered that none of all this has any relevance 
unless it is somehow shown that the solutions of the time-dependent 
problem corresponding to the original Smoluchowski equations (\ref{eq:1.2}) 
can be shown to approach the scaling limit. It is possible that the 
recursive structure underlying the diagonal kernel might make this result 
somewhat easier to show. It remains, however, a formidable task, left to 
future research. 
\section*{Acknowledgements}
It is a pleasure to acknowledge helpful conversations with H.~Larralde, as 
well as Ph.~Lauren\c cot and M.~Escobedo, as well as very useful remarks 
from the referee. Most of all, however, I would 
like to acknowledge Francesco Calogero's contribution: through his 
persistently critical attitude to scaling theory expressed in many 
discussions, 
he has forced me to look 
at some of the issues treated here with partial success: 
this paper would not have been begun without his comments. It is 
therefore gratefully dedicated to him on the occasion of his birthday.

\label{lastpage}

\end{document}